\documentstyle[aps,eqsecnum]{revtex}

\begin{document}

 \draft

\title{\bf Light meson masses and mixings}

\author{Loyal Durand\thanks{Electronic address: 
ldurand@theory1.hep.wisc.edu}}
\address{ Department of Physics, University of Wisconsin-Madison,
Madison, WI 53706}

\date{\today}
\maketitle

\begin{abstract}
I present a simple discussion of the masses and mixings of the light pseudoscalar and vector mesons based on a ``$q\bar{q}$'' description of the effective field theory. The analysis includes $\eta'$(958) from the beginning, and is largely concerned with structural questions. While the final results are mostly known, the method gives insight into the general form of the meson mass matrices and the different character of the mass splittings and mixings in the pseudoscalar and vector multiplets, and provides a coherent overall picture
\end{abstract}


\section{INTRODUCTION}
\label{sec:introduction}

The theoretical analysis of the masses and mixings of the light pseudoscalar and vector mesons has a long history, from the original Gell-Mann--Okubo mass formula \cite{Gell-Mann,Okubo} through modern analyses based on chiral perturbation theory as summarized, for example, in \cite{Gasser,DGH}. The masses of the light pseudoscalar mesons are now discussed in terms of the breaking of the chiral ${\rm SU(3)}_L\times{\rm SU(3)}_R$ symmetry that exists in QCD for massless quarks, with the octet mesons appearing initially as the eight Goldstone bosons for the broken symmetry. The SU(3) singlet state $\left|\eta_1\right>\approx\left|\eta'\right>$ and the vector mesons are treated separately. 

A perturbative treatment of the effect of quark masses on the pseudoscalar mesons, introduced through the quark mass matrix
\begin{equation}
\label{quark_masses}
\left(\begin{array}{ccc}
m_u & 0 & 0 \\
0 & m_d & 0\\
0 & 0 & m_s
\end{array}
\right)
\end{equation}
leads to the prediction that \cite{Gasser,DGH}
\begin{eqnarray}
\label{perturbative_masses}
m_\pi^2=2B_0\hat{m}, \quad && m_K^2=(m_s+\hat{m})B_0, \nonumber \\
m_{\eta_8}^2=\frac{2}{3}(2m_s+\hat{m})B_0, \quad && m_{\eta_1}^2 = m_0^2+\frac{2}{3}(m_s+2\hat{m})B_0.
\end{eqnarray}
Here $\hat{m}=(m_u+m_d)/2$ is the average light-quark mass, $m_K^2=(m_{K^-}^2+m_{K^0}^2)/2$, and $m_{\eta_8}$ and $m_{\eta_1}$ are the masses of the unmixed SU(3) isospin-0 octet and singlet states $\left|\eta_8\right>$ and $\left|\eta_1\right>$, $B_0$ is an unknown dynamical matrix element, and $m_0$ is the original mass of the $\eta_1$, also unknown. Finally, the original states  $\left|\eta_8\right>$ and $\left|\eta_1\right>$ couple through an off diagonal matrix element 
\begin{equation}
\label{m^2_8,1}
m_{8,1}^2=-\frac{2\sqrt{2}}{3}(m_s-\hat{m})B_0.
\end{equation}
to produce the observed mass eigenstates  $\left|\eta\right> \equiv \eta$(547) and $\left|\eta'\right>\equiv\eta$(958) \cite{PDG}. 

The perturbative predictions are reasonably successful. Equation (\ref{perturbative_masses}) contains the Gell-Mann--Okubo mass relation 
\begin{equation}
m_{\eta_8}^2 =  \frac{1}{3}\left(4M_K^2-m_\pi^2\right),
\end{equation}
which is valid to about 20\% when the $\eta_8$ is identified with the $\eta$. The $\eta$-$\eta'$ mixing angle defined by
\begin{equation}
\label{def:mixing_angle}
\left|\eta'\right>=\sin\theta_P \left|\eta_8\right> + \cos\theta_P\left|\eta_1\right>  , \quad \left|\eta\right>=\cos\theta_P \left|\eta_8\right> - \sin\theta_P\left|\eta_1\right> ,
\end{equation}
is predicted to be $-10^\circ$ \cite{PDG}, at the edge of the range $-13^\circ$ to $22^\circ$ obtained in various fits to decay data, for example, in
\cite{DGH,DHL1985,Gilman,Bramon1990,Bramon1999,Venugopal}. 

The situation is different for the vector mesons, where the chiral symmetry arguments do not apply. However, symmetry ideas from SU(6) and quark-model and similar methods again lead to successful mass relations among the vector mesons, while dynamical ideas related to annihilation processes have been invoked to explain the different mixing patterns \cite{Isgur,DGH}.

In the present paper, I present a simple structural analysis of the mass and mixing problems for the nonets of pseudoscalar and vector mesons. Most of the ideas appear in the literature as applied to specific problems, and most of the results are also familiar. However, by introducing some quark-mass effects which modify previous approaches and lead naturally to mass matrices of the most general type \cite{Morpurgo}, I find a consistent overall picture in which, for example, the different mass and mixing patterns for the pseudoscalar and vector mesons follow from physical input, and the ratios of dynamical matrix elements in the two multiplets give information on the mixings. I first treat the pseudoscalar nonet, with the $\eta'$ included from the outset. I then treat the vector nonet, and comment on the connections and differences, and conclude with a few overall remarks.

\section{THE PSEUDOSCALARS}
\label{sec:pseudoscalars}

\subsection{Structure of the mass matrix to O($m_q$)}
\label{subsec:structure}

In this section, I discuss a generalized version of the mass matrix which allows actual fits to the masses of the pseudoscalar mesons. It will be useful to start in a ``quark model'' basis of effective meson fields $\phi_{ij}(x)$ and meson states $\left| \phi_{ij}\right>$, where $i,\,j$ label the flavor structure in a ${\bf 3}\times\bar{\bf 3}$ description of a flavor SU(3). The general basis includes both the SU(3) octet and singlet states. The mass operator for the pseudoscalar mesons is then of the form
\begin{equation}
\label{M^2_op}
M^{\,2}_{op}=\sum_{i'j';ij}\phi^\dagger_{i'j'}M^{\,2}_{i'j';ij}\phi_{ij}.
\end{equation}
I will work throughout with the matrix $M^2$, and will first consider the neutral sector. 

The combinations $u\bar{u}$, $d\bar{d}$, and $s\bar{s}$ for the flavor labels
$i,\,j$ are completely equivalent for massless quarks in QCD, so, with a suitable choice of phases, the initial mass matrix for the neutral states must have the same entry in every location corresponding to an SU(3) singlet configuration,
\begin{equation}
\label{M_0^2}
M_{0,P}^{\,2} =\frac{1}{3}\left(
\begin{array}{ccc}
m_0^2 & m_0^2 & m_0^2 \\
m_0^2 & m_0^2 & m_0^2 \\
m_0^2 & m_0^2 & m_0^2 
\end{array}\right)
= B_0\left(
\begin{array}{ccc}
m_1 & m_1 & m_1 \\
m_1 & m_1 & m_1 \\
m_1 & m_1 & m_1 
\end{array}\right)\,.
\end{equation}
The factor $B_0$ in the second form has been extracted for convenience in writing later equations. $M_{0,P}^{\,2}$ has eigenvalues zero in the SU(3) octet states $\left|\pi\right>$ and $\left|\eta_8\right>$, or any combination thereof, and the eigenvalue $m_0^2=3B_0m_1$ for the SU(3) singlet $\left|\eta_1\right>$. The pion and $\eta_8$ are therefore massless, as expected for Goldstone bosons. The term $m_0^2$ contains contributions from the axial anomaly \cite{DGH} and from annihilation diagrams such as that in Fig.\ \ref{fig1}\,(a). Both lead to the singlet structure in Eq.\ (\ref{M_0^2}) when written in the $q\bar{q}$ basis. 

The structure changes in the presence of quark masses, both through the addition of diagonal contributions proportional to the quark masses, and through the effect of those masses on the annihilation terms. The effect of the latter is neglected in all discussions of which I am aware. The modified mass matrix has an obvious structure 
\begin{equation}
\label{M^2_complex}
M_P^{\,2} = B_0\left(
\begin{array}{ccc}
m_1+2m_u +2am_u & m_1 + a(m_u+m_d) & m_1 + a(m_u+m_s)\\
m_1+a(m_d+m_u) & m_1+2m_d +2am_d & m_1+a(m_d+m_s) \\
m_1+a(m_s+m_u) & m_1+a(m_s+m_d) & m_1+2m_s+2am_s
\end{array}\right)
\end{equation}
in the $q\bar{q}$ representation, where the terms proportional to $a$ arise from mass insertions in the annihilation diagrams.\footnote{The basic structure of $M_P^{\,2}$ was apparently first suggested by Isgur \cite{Isgur}, but without the $a$ terms. Isgur identified what are here the $B_0m_1$ terms with the annihilation diagrams, treated the diagonal terms phenomenologically as effective quark masses, and used the structure to explain the difference between the $\eta,\,\eta'$ and $\omega,\,\phi$ mixings. The general form of the mass matrix was  given by Morpurgo \cite{Morpurgo}, but without the physical identification of the various contributions. } I expect $a$ to be small, with $|a|m_s \ll m_1$. A dispersion relation argument shows that the annihilation contribution to $m_1$ is positive in the symmetrical limit. Its magnitude is decreased by mass insertions in the loop, so $a$ should be negative and small.\footnote{The $\eta'$ acquires mass through the $m_1$ terms without any quark mass insertions. As a result, the off-diagonal contributions proportional to $m_s$ can be complex, with $M_P^{\,2}$ still Hermitian. The imaginary part of $a$ should be small, and can be absorbed to first order in a change in the phase of $\left|\eta_1\right>$. I will therefore take $a$ as real.} The light quark masses are also small, $m_u,\, m_d\ll m_s$.  

A transformation to the SU(3) basis gives a considerably less transparent structure. The resulting matrix can be simplified by dropping a diagonal term proportional to $a\hat{m}$ and off-diagonal terms which involve $m_u-m_d$. Both are small, with $|a\hat{m}|,\,|m_u-m_d|\ll m_s,\,m_1$. The off-diagonal terms induce a small mixing of the $\pi^0$ with the $\eta$ and $\eta'$, but first contribute to the physical masses in second order \cite{Gasser}. The reduced matrix is
\begin{equation}
\label{M^2}
M_P^{\,2} = B_0\left(
\begin{array}{ccc}
2\hat{m} & 0 & 0 \\
0 & 2\hat{m}+4\hat{m}_s/3 & \ -\sqrt{2}(2+3a)\hat{m}_s/3  \\
0 &  -\sqrt{2}(2+3a)\hat{m}_s/3  & \ 2\hat{m} + 3m_1 + 2(1+3a)\hat{m}_s/3
\end{array}\right)\,,
\end{equation}
where $\hat{m}=(m_u+m_d)/2$ and $\hat{m}_s=m_s-\hat{m}$. It is convenient to rewrite this as 
\begin{equation}
\label{M=1+Mhat}
M_P^{\,2}=m_\pi^2\openone + \left(
\begin{array}{cc}
0 & {\bf 0} \\
{\bf 0}^T &\hat{M}_P^{\,2}
\end{array}
\right)\,,
\end{equation}
where $\openone$ is the 3$\times$3 unit matrix, $\bf 0$ is the 1$\times$2 null matrix, and $\hat{M}_P^{\,2}$ is the $2\times$2 matrix
\begin{equation}
\label{M^2_reduced}
\hat{M}_P^{\,2}=B_0\left(
\begin{array}{cc}
 4\hat{m}_s/3 & -\sqrt{2}(2+3a)\hat{m}_s/3 \\
 -\sqrt{2}(2+3a)\hat{m}_s/3 & 3m_1 + 2(1+3a)\hat{m}_s/3
\end{array} \right)\,.
\end{equation}
The matrix $m_\pi^2\openone$ contributes a term $m_\pi^2$ to each of the eigenvalues of $M_P^2$. I will drop this term and work with the 2$\times$2 matrix $\hat{M}_P^{\,2}$ which has eigenvalues $\hat{m}_\eta^2\equiv m_\eta^2-m_\pi^2$ and $\hat{m}_{\eta'}^2\equiv m_{\eta'}^2-m_\pi^2$. 

$\hat{M}_P^{\,2}$ has the form of the general $\eta_8,\,\eta_1$ mass matrix \cite{PDG,Morpurgo},
\begin{equation}
\label{M_18}
M_{\eta\eta'}^{\,2}=\left(
\begin{array}{ccc}
M_{88}^2 & M_{18}^2 \\
M_{18}^2 & M_{11}^2
\end{array}\right)\,,
\end{equation}
with $M_{18}^2$ real after a phase has been absorbed. In particular, both involve three independent parameters. I find it striking that the explicit inclusion of the annihilation terms already leads to this structure at first order in $\hat{m}_s$. 

The kaon and charged pion masses have the usual simple structure which does not involve annihilation contributions, with
\begin{equation}
\label{m_K,m_pi}
m_{K^+}^2=B_0(m_s+m_u), \qquad  m_{K^0}^2=B_0(m_s+m_d), \qquad m_{\pi^\pm}^2 = B_0(m_u+m_d)=2B_0\hat{m}.
\end{equation}
The average kaon mass is $m_K^2\equiv(m_{K^+}^2+m_{K^-}^2)/2=B_0(m_u+\hat{m})$, so
\begin{equation}
\label{m_K}
m_K^2-m_\pi^2\equiv\hat{m}_K^2=B_0\hat{m}_s.
\end{equation}

\subsection{Physical masses and fits at O($m_q$)}
\label{subsec:fits}

The masses of the physical particles $\eta$ and $\eta'$ are determined by the eigenvalues of $\hat{M}_P^{\,2}$,  
\renewcommand{\arraystretch}{1.2}
\begin{eqnarray}
\label{eta_mass}
\hat{m}_\eta^2  &=&  \frac{1}{2}B_0\left[3m_1+2(1+a)m_s\right] - \frac{1}{2}B_0\left[(3m_1+2(1+a)m_s)^2 - 8m_s(2m_1-a^2m_s)\right]^{1/2} \nonumber\\ 
&\mbox{$\approx$}& \frac{4}{3}B_0m_s\left[1 - \frac{2}{9}\left(1 + \frac{3}{2}a\right)^2 \frac{m_s}{m_1} +\cdots\right], 
\\
\label{eta'_mass}
\hat{m}_{\eta'}^2 &=& \frac{1}{2}B_0\left[3m_1+2(1+a)m_s\right] + \frac{1}{2}B_0\left[(3m_1+2(1+a)m_s)^2 - 8m_s(2m_1-a^2m_s)\right]^{1/2}
\nonumber \\
&\mbox{$\approx$}& 3B_0m_1+\frac{2}{3}B_0m_s\left[1+3a + \frac{4}{9}\left(1 + \frac{3}{2}a\right)^2 \frac{m_s}{m_1} +\cdots \right], 
\end{eqnarray}
\renewcommand{\arraystretch}{1.0}
where $\hat{m}_{\eta'}^2 = m_{\eta'}^2-m_\pi^2$ and $\hat{m}_\eta^2= m_\eta^2-m_\pi^2$. The leading terms give the usual first-order perturbative result except for the independent appearance of $a$ in the expressions for $\hat{m}_{\eta'}^2$ and the $\eta,\,\eta'$ mixing angle.  

If $a$ is small, we obtain the inverse relations
\begin{equation}
\label{estimates}
B_0m_s\approx\frac{3}{4}\hat{m}_\eta^2, \qquad B_0m_1\approx\frac{1}{3} \left(\hat{m}_{\eta'}^2 - \frac{1}{2}\hat{m}_\eta^2\right),
\end{equation}
for $B_0m_s$ and $B_0m_1$. The first becomes the Gell-Mann--Okubo relation  if $B_0m_s$ is identified with $\hat{m}_K^2$ as in Eq.\ (\ref{m_K}).  These relations give the numerical estimates $B_0m_s\approx 0.211$ GeV$^2$, $B_0m_1\approx 0.252$ GeV$^2$, and $m_s/m_1\approx 0.873$ for the physical masses $\hat{m}_{\eta'}^2=0.898$ GeV$^2$, $\hat{m}_\eta^2 = 0.281$ GeV$^2$. The expansion parameter $2m_s/9m_1\approx 0.19$ in Eqs.\ (\ref{eta_mass}) and (\ref{eta'_mass}) is therefore small, but not so small that the correction terms can be ignored. Since the ``first order'' fit to the observed masses automatically incorporates these corrections and other diagonal $m_s^2$ terms in the fitted values of $m_1$ and $M_s$, it seems preferable to use the general form of the eigenvalues  or $\hat{M}_P^{\,2}$ in physical analyses.

The eigenvalues of the general mass matrix $\hat{M}_P^{\,2}$ are determined by the conditions
\begin{eqnarray}
\label{trace_relations}
\hat{m}_{\eta'}^2+\hat{m}_\eta^2 &=& {\rm Tr}\,\hat{M}_P^{\,2} = B_0[3m_1+2(1+a)m_s]\\
\label{det_relation}
\hat{m}_{\eta'}^2\hat{m}_\eta^2 &=& {\rm det}\,\hat{M}_P^{\,2} = B_0^2(4m_1m_s-2a^2m_s^2).
\end{eqnarray}
The $a^2m_s^2$ term in the last expression is presumably small, and will be dropped in the immediately following equations for simplicity.\footnote{It will actually be small for the final range of parameters discussed below. Including this term pushes the limit on $a$ in Eq.\ (\ref{a_limit}) farther from zero.}

Eliminating $m_s$ between the two equations and solving for $m_1$ gives the relation
\begin{equation}
\label{m1}
B_0 m_1=\frac{1}{6}\left(\hat{m}_{\eta'}^2+\hat{m}_\eta^2\right) + \frac{1}{6} \left[\left(\hat{m}_{\eta'}^2+\hat{m}_\eta^2\right)^2-6\hat{m}_{\eta'}^2 \hat{m}_\eta^2\left(1+a\right)\right]^{1/2}.
\end{equation}
The argument of the radical is negative for the physical masses, and there is no real solution for $m_1$, unless
\begin{equation}
\label{a_limit} 
a< \left[\left(\hat{m}_{\eta'}^2+\hat{m}_\eta^2\right)^2-6\hat{m}_{\eta'}^2 \hat{m}_\eta^2\right]/6\hat{m}_{\eta'}^2 \hat{m}_\eta^2=-0.084.
\end{equation}
This gives a useful constraint on our general parametrization of $\hat{M}_P^{\,2}$  and requires $a$ to be negative, as expected.    

Once a value of $a$ is given, Eq.\ (\ref{m1}) determines $B_0m_1$, and $B_0m_s$ then follows from Eq.\ (\ref{det_relation}),
\begin{equation}
\label{ms}
B_0m_s=\hat{m}_{\eta'}^2\hat{m}_\eta^2/4B_0m_1 ,
\end{equation}
with possible corrections of relative order $a^2m_s/2m_1$.

We need one further condition to determine $a$. This is given within the $\eta_1,\,\eta_8$ sector by the mixing angle $\theta_P$ defined in Eq.\ (\ref{def:mixing_angle}),
\begin{eqnarray}
\label{tantheta}
\tan\theta_P &=& -2\sqrt{2}\left(2+3a\right)m_s/ 
\left\{9m_1-2\left(1-3a\right)m_s \right. \nonumber \\ && \left. + 3\left[ \left(3m_1+2m_s+2am_s\right)^2 - 8\left(2m_1-a^2m_s\right)m_s\right]^{1/2}\right\}.
\end{eqnarray}
This angle is not well determined experimentally. Different analyses give values of $-13^\circ$ to $-22^\circ$ depending on the data used\cite{DGH,DHL1985,Gilman,Bramon1990,Bramon1999,Venugopal}, and the theoretical connections assumed to hold between decay matrix elements and the meson masses and mixings. 

A different approach is to take the value of $B_0m_s\approx \hat{m}_K^2$ as known. The parameters $m_1$ and $a$ are then determined by the $\eta$ and $\eta'$ masses, and $\theta_P$ can be predicted. I will adopt this approach, but will allow for some difference between the effective values of $B_0m_s$ in the kaon and $\eta,\,\eta'$ systems by introducing a multiplicative correction factor $x$, with $B_0m_s=x\hat{m}_K^2$. This difference can arise from higher order effects in the chiral expansion, loop corrections, or the change in matrix elements expected in the change from a system with one heavy quark to systems which can have two. All of these corrections can be encompassed in our general parametrization of $\hat{M}_P^{\,2}$; their separation depends on dynamical calculation beyond the scope of this paper. The net effect gives $x\not=1$, with the expectation that $x-1$ is on the order of the fractional violation of the Gell-Mann--Okubo relation.

The results obtained in our approach are summarized in Figs.\ \ref{fig2} and \ref{fig3} which show the fitted value of $a$ and the predicted mixing angle $\theta_P$ as functions of $x$. It appears from the results on mixing angles that $x$ is in the range $1.0\lesssim x \lesssim 1.2$, a reasonable result, with corresponding values of $a$ in the range $-0.24\lesssim a\lesssim -0.1$.  The parameter $a$ is therefore small, as expected. An argument given later in connection with Eq.\ (\ref{b0/b'0}) suggests that, when the vector mesons are included in the analysis, values $x\approx 1.05$, $a\approx -0.18$,  and $\theta_P\approx-15^\circ$ are favored. This predicted mixing is in the center of the ranges found in various analyses \cite{DHL1985,Gilman,Bramon1990,Bramon1999,Venugopal}.

\section{THE VECTORS}
\label{sec:vectors}

The situation for the vector mesons $\rho$, $\omega$, and $\phi$ is quite different. In the $q\bar{q}$ basis, the initial mass matrix has the form of the pseudoscalar mass matrix in Eq.\ (\ref{M^2_complex}) with the addition of a diagonal term $B'_0m_V\openone$ which gives a common mass to all the vector states. No such term was not allowed in the pseudoscalar sector by the requirement that the pion and $\eta_8$ appear initially as Goldstone bosons of the broken $SU_L(3)\times SU_R(3)$ symmetry. Furthermore, the elements analogous to the terms $B_0m_1$ in Eqs.\ (\ref{M_0^2}) or (\ref{M^2_complex})  have no anomalous component, so obtain contributions only from annihilation diagrams. The leading vector annihilation diagram involves three-gluon exchange as shown in Fig.\ \ref{fig1}\,(b). The corresponding matrix element, which I will denote by $B'_0m_2$, is expected to be much smaller than the two-gluon contribution to the pseudoscalar masses from Fig.\ \ref{fig1}\,(a), but still positive with $m_2\ll m_1$. The relation $m_1\simeq m_s$ from the pseudoscalar sector then implies that $m_2\ll m_s$. The effect of mass insertions on the already-small $m_2$ terms should also be small, with $|a'|\ll 1$. Dropping those corrections, we obtain the simplified vector mass matrix 
\begin{equation}
\label{M_V}
\hat{M}_V^{\,2}= B'_0\left(
\begin{array}{ccc}
m_V+m_2+2m_u & m_2 & m_2 \\
m_2 & m_V+m_2+2m_d & m_2 \\
m_2 &m_2 & m_V+ m_2+2m_s 
\end{array}\right)\,.
\end{equation}
The matrix elements $B_0m_s$ and $B'_0m_s$ differ in the dynamical quark model only by the effects of spin-dependent interactions on the wave functions and are expected to be similar in magnitude, $B_0\simeq B'_0$.

In contrast to $M_P^{\,2}$, which was dominated by the flavor-singlet structure of the $m_1$ terms in Eq.\ (\ref{M_0^2}) with $2m_s< 3m_1$ appearing as a perturbation, $M_V^{\,2}$ is dominated by the the strange-quark mass term. There is consequently no reason to transform to the SU(3) basis used to pick out the singlet structure. The $q\bar{q}$ basis is natural. It is useful, however, to transform to a flavor SU(2) basis $\left|\phi_1\right>=\left(\left|\phi_{u\bar{u}}\right> - \left|\phi_{d\bar{d}}\right>\right)/\sqrt{2}$, $\left|\phi_0\right>=\left(\left|\phi_{u\bar{u}}\right> + \left|\phi_{d\bar{d}}\right>\right)/\sqrt{2}$ for the light-quark sector. This isolates the isovector component $\left|\phi_1\right>=\left|\rho\right>$, and brings the mass matrix to the form
\begin{equation}
\label{M_V+hatM_V}
M_V^{\,2}=B'_0(m_v+2\hat{m})\openone+\left(\begin{array}{cc}
0 & {\bf 0} \\
{\bf 0}^T &\hat{M}_V^{\,2}
\end{array}
\right)\,,
\end{equation}
where I have dropped off-diagonal terms in $m_u-m_d$ which contribute linearly to the $\rho,\,\omega$ mixing \cite{Gasser}, but only quadratically to the masses. The matrix $\hat{M}_V^{\,2}$ is
\begin{equation}
\label{hatM_V}
\hat{M}_V^{\,2}=B'_0\left(\begin{array}{cc}
2m_2 & \sqrt{2}m_2 \\
\sqrt{2}m_2 & \ 2\hat{m}_s+m_2 
\end{array} \right),
\end{equation}
where $\hat{m}_s=m_s-\hat{m}$ as before.

The mass of the $\rho$ meson is clearly given by $m_\rho^2=B'_0(m_v+2\hat{m})$. The eigenvalues of $\hat{M}_V^{\,2}$, taken to first order in the small quantity $m_2/m_s$, are
\begin{equation}
\label{Vmasses}
 \hat{m}_\omega^2=m_\omega^2-m_\rho^2=2B'_0 m_2 \quad {\rm and}\quad \hat{m}_\phi^2=m_\phi^2-m_\rho^2=B'_0\left( 2m_s+m_2 \right).
\end{equation}
These relations give $B'_0(m_V+2\hat{m})=0.592$ GeV$^2$, $2B'_0m_2=0.0206$ GeV$^2$, and $2B'_0m_s=0.437$ GeV$^2$. Thus, $m_2/m_s\approx 0.047$, and $m_2\ll m_s$ as expected. The approximate degeneracy of the $\rho$ and $\omega$ masses follows from the smallness of $m_2$. This is expected in QCD as argued above. There is no constraint on the extra parameter $a'$ in the general case since $\hat{m}_\omega^2\hat{m}_\phi^2\ll(\hat{m}_\omega^2+\hat{m}_\phi^2)^2$.

The average $K^*$ mass squared $\hat{m}_{K^*}^2$ is defined by
\begin{equation}
\label{m_K*}
\hat{m}_{K^*}^2=\left(m_{K^+}^2+m_{K^0}^2\right)/2-m_\rho^2 = B'_0m_s.
\end{equation}
The combination of the two preceding equations gives the sum rule
\begin{equation}
\label{sum_rule}
\hat{m}_\phi^2-\frac{1}{2}\hat{m}_\omega^2=2\hat{m}_{K^*}^2.
\end{equation}
This is satisfied to 6\% ($0.437\ {\rm{GeV}^2\approx 0.412\ \rm GeV}^2$). However, this sum rule is subject to some uncertainty theoretically because of different effect of the O($m_s^2$) contributions and loop corrections on the $\omega,\,\phi$ and $K^*$ systems. If the uncertainties are incorporated in a scaling of the $K^*$ contribution with $B'_0m_s=x'\hat{m}_{K^*}^2$,  the scale factor $x'$  is 1.06, in the favored range for the scale factor $x$ for the pseudoscalar mesons.

A comparison of Eqs.\ (\ref{estimates}) and (\ref{Vmasses}) leads to the rough estimate $B_0/B'_0\approx 3\hat{m}_\eta^2/(2\hat{m}_\phi^2-\hat{m}_\omega^2) \approx 0.96$ while the ratio $\hat{m}_K^2/\hat{m}_{K^*}^2$ gives $B_0/B'_0\approx 1.10$. With either estimate, $B_0/B'_0\approx 1$ as expected. If I include the leading correction to $\hat{m}_\eta^2$ from Eq.\ (\ref{eta_mass}), I find the more precise relation
\begin{equation}
\label{b0/b'0}
\frac{B_0}{B'_0}=\frac{x\hat{m}_K^2}{x'\hat{m}_{K^*}^2} \approx\frac{3\hat{m}_\eta^2}{2\hat{m}_\phi^2-\hat{m}_\omega^2} \left[1-\frac{2}{9}\left(1+\frac{3}{2}a\right)^2\frac{m_s}{m_1}+ \cdots\right]^{-1}.
\end{equation}
This is satisfied for $x=1.05$, $a=-0.18$. The value of $x$ is in the range $1\lesssim x\lesssim 1.2$ consistent with analyses of $\eta,\,\eta'$ mixing, and gives the value $\theta_P=-15^\circ$. The whole picture is therefore consistent.

I will define the mixing angle $\theta_V$ connecting the isospin zero state $\left|\phi_0\right>$ with $\left|\phi_{s\bar{s}}\right>$ so that
\begin{equation}
\label{vector_mixing}
\left|\omega\right>=\cos\theta_V\left|\phi_0\right> - \sin\theta_V \left|\phi_{s\bar{s}}\right> , \quad \left|\phi\right>=\sin\theta_V\left|\phi_0\right> + \cos\theta_V \left|\phi_{s\bar{s}}\right>.
\end{equation}
The results above give
\begin{equation}
\label{vec_mixing}
\theta_V\approx\frac{m_2}{\sqrt{2}m_s}\left(1+\frac{m_2}{2m_s}\cdots\right).
\end{equation}
Numerically, $\theta_V\approx m_2/\sqrt{2}m_s \approx 1.95^\circ $, so there is very little mixing. The angle $\theta'_V$ defined with respect to the SU(3) basis $\left|\omega_8\right>$,  $\left|\omega_1\right>$ \cite{PDG} is related to $\theta_V$ by
\begin{equation}
\label{theta'_V}
\theta'_V=\theta_V+\arctan(1/\sqrt{2}) = \theta_V+35.3^\circ,
\end{equation}
where $\arctan(1/\sqrt{2})=35.3^\circ$ is the angle for so-called ``ideal'' SU(3) mixing. The sum gives $\theta'_V\approx 37.2^\circ$, in good agreement with experiment. However, I find little reason to use the SU(3) description. The inequality $m_s \gg m_2$ picks out the ``$q\bar{q}$'' basis states $\left|\phi_{ij}\right>$ as giving the natural description of the problem, with  the $\rho$ split off in the isospin triplet substate. The physical masses then correspond yield a small mixing angle $\theta_V\approx 1.9^\circ$ without special assumptions.

It is interesting to compare this result to that for the pseudoscalars. In the latter case, $m_s\simeq m_1$, so the SU(3) singlet is not clearly singled out, and it is less certain what basis is to be preferred. $\hat{M}_P^{\,2}$ is given in the $\left|\phi_0\right>$, $\left|\phi_{s\bar{s}}\right>$ basis by
\begin{equation}
\label{M_P_ssbar}
\hat{M}_P^{\,2} = \left( \begin{array}{cc}
2m_1 & \sqrt{2}(m_1+am_s) \\
\sqrt{2}(m_1+am_s) &\  m_1+2(1+a)m_s
\end{array} \right)\,.
\end{equation}
The $\phi_0,\,\phi_{s\bar{s}}$ mixing angle $\theta'_P$ defined by the relations
\begin{equation}
\left|\eta'\right>=\sin\theta'_P \left|\phi_0\right> + \cos\theta_P\left|\phi_{s\bar{s}}\right>  , \quad \left|\eta\right>=\cos\theta_P \left|\phi_0\right> - \sin\theta_P\left|\phi_{s\bar{s}}\right> ,
\end{equation}
is given by
\begin{eqnarray}
\label{theta'_P}
\tan\theta'_P &=& 2\sqrt{2}\left(m_1+am_s\right)/\left\{ -m_1+2\left(1+a\right)m_s \right. \nonumber \\
&& \left. + \left[ \left(3m_1+2m_s+2am_s\right)^2 - 8\left(2m_1-a^2m_s\right)m_s\right]^{1/2}\right\} \\
&\approx& \sqrt{2}\left(m_1+am_s\right)/\left[m_1+2(1+a)m_s\right].
\end{eqnarray}
This gives a rather large mixing angle in the range 47$^\circ$ to 33$^\circ$ for $x$ from 0.9 to 1.2, a consequence of the similar sizes of $m_1$ and $m_s$.
The SU(3) basis is therefore somewhat better physically. In contrast, given the result $m_s\simeq m_1$ for the pseudoscalars, there is a clear physical argument that $m_2\ll m_1\simeq m_s$, so that the $\phi_0,\,\phi_{s\bar{s}}$ basis is preferred for the vectors, with the mixing  small corresponding to ``ideal'' mixing in the SU(3) basis.   

\section{CONCLUSIONS}
\label{sec:conclusions}

\begin{figure}
\centerline{
\epsffile{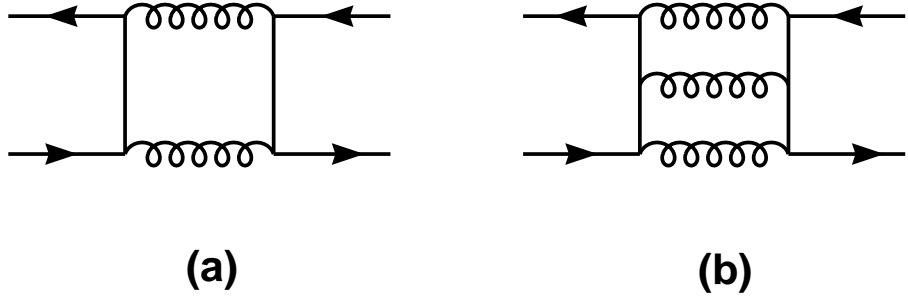}
}
\caption{Annihilation diagrams which contribute to the symmetric mass matrix $M_0^2$ for (a), the pseudoscalar mesons, and (b), the vector mesons.
}
\label{fig1}
\end{figure}

\begin{figure}
\def\picsize{6in}                  
\def\picfilename{fig2.eps}         
\centerline{\epsfxsize\picsize\epsfbox{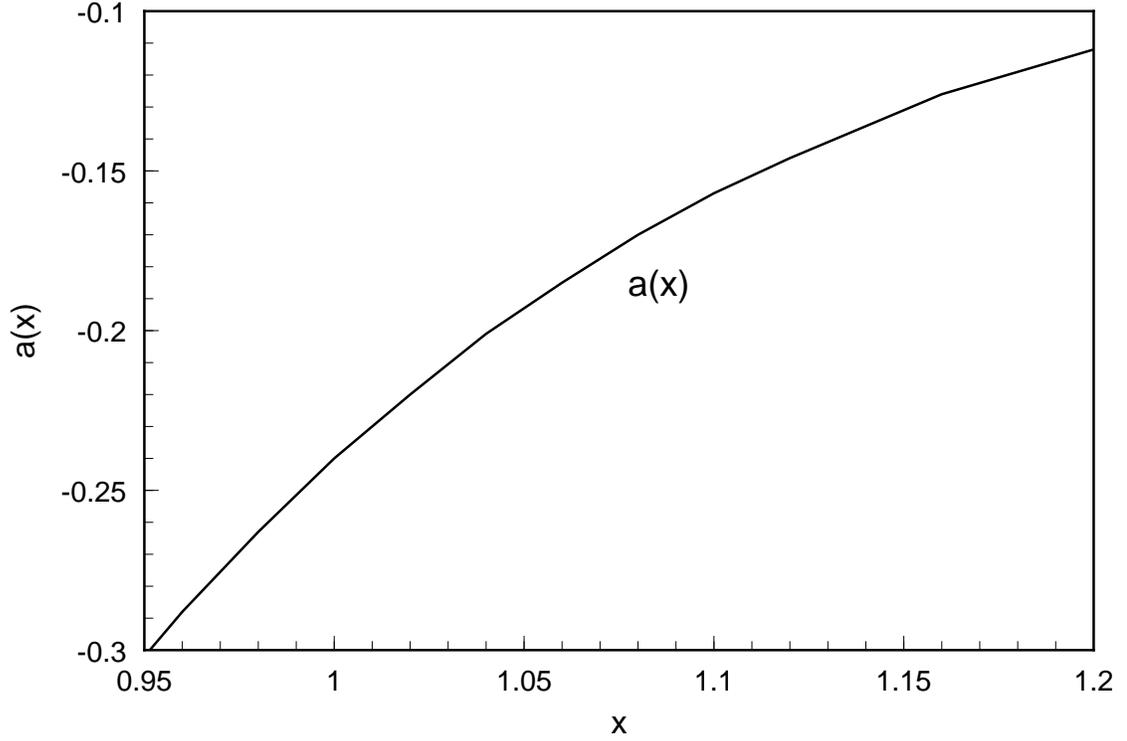}}      
\caption{Variation of the parameter $a$ in the mass matrix $\hat{M}_P^{\,2}$ with the scale factor $x$ in the relation $B_0m_s=x\hat{m}_K^2$. Values of $x$ in the range $1.05\protect\lesssim x\protect\lesssim 1.2$ are generally favored by determinations of the mixing angle $\theta_P$ from decay data.
}
\label{fig2}
\end{figure}

\begin{figure}
\def\picsize{6in}                  
\def\picfilename{fig3.eps}         
\centerline{\epsfxsize\picsize\epsfbox{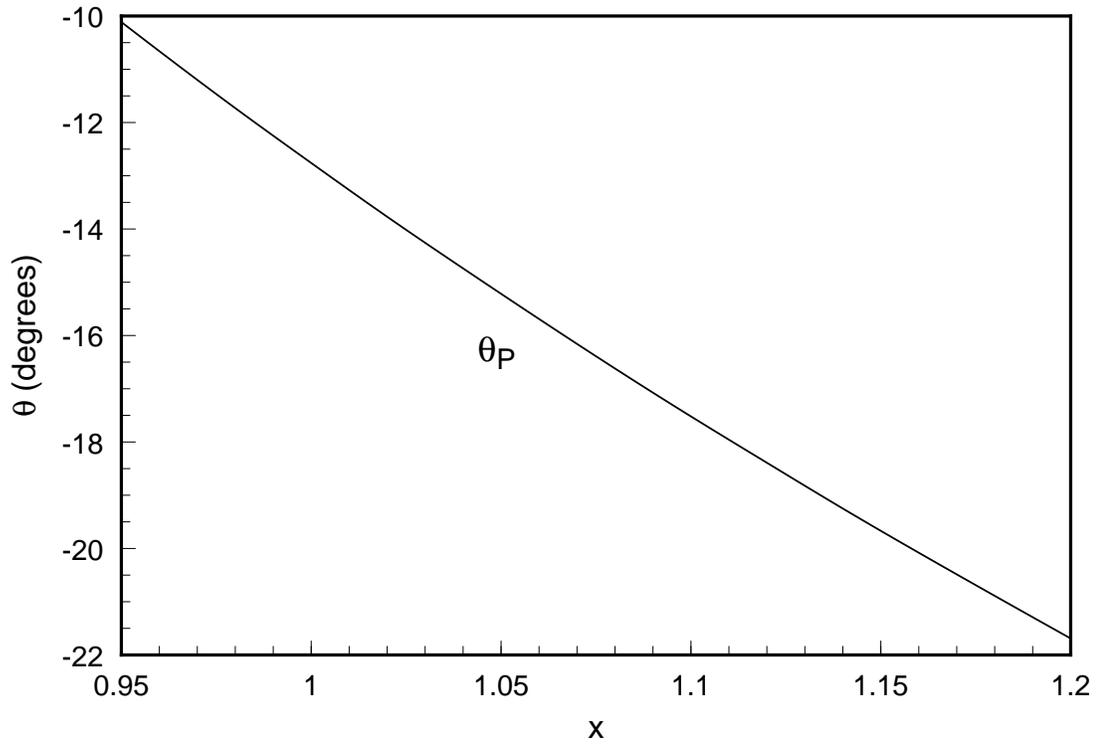}}      
\caption{Dependence of the $\eta_8,\,\eta_1$ mixing angle $\theta_P$, Eq.\ (\protect\ref{tantheta}), on the scale factor $x$ in the relation $B_0m_s=x\hat{m}_K^2$. Values of $\theta_P$ in the range $-15^\circ$ to $-20^\circ$, or $1.05\protect\lesssim x\protect\lesssim 1.2$ are generally preferred from analyses of meson decay data.
}
\label{fig3}
\end{figure}

My conclusions are simple: there is a substantial advantage in treating the  nonets of pseudoscalar and vector meson together in analyzing the patterns of masses and mixings. It is important, in particular, to include the $\eta'(958)$ in the analysis to get a consistent overall picture. Most of the results presented here are well-known, though not treated as a whole in other discussions. However, the complete analysis given here introduces a new feature through the through the inclusion of the leading quark mass corrections to the annihilation diagrams in Fig. \ref{fig1}. This leads to the general mass matrix for the pseudoscalar  mesons in Eq.\ (\ref{M^2_complex}), and a similar result for the vector mesons. These reduce to the simpler forms in Eqs.\ (\ref{M=1+Mhat}) and (\ref{M_V+hatM_V}) if small off-diagonal terms proportional to $m_u-m_d$ are neglected. The residual 2$\times$2 mass matrices have a completely general form.

I show in the case of the pseudoscalar mesons that the physical $\eta$ and $\eta'$ masses can only be fitted if the annihilation parameter $a$ in the mass matrix is nonzero and negative. The wide splitting of the $\pi$, $\eta$, and $\eta'$ masses, and fairly large mixing angle $\theta_P$ for the pseudoscalars, reflect the fact that the strange-quark mass insertion $B_0m_s$ and the basic SU(3) singlet term $B_0m_1$ are similar in magnitude.

The situation is much simpler for the vector mesons. While the initial mass matrix has the same formal structure as that for the pseudoscalar mesons, the large anomalous contribution to the  singlet component of the latter is missing, and the annihilation diagram is suppressed by the necessity of having at least three gluons in the intermediate state. As a result, the parameter $m_2$ analogous to $m_1$ is expected to be small in QCD, with $m_2\ll m_1 \approx m_s$, and the mass matrix simplifies. The smallness of $m_2$ leads to the near-degeneracy of the $\rho$ and $\omega$ masses, and to the small mixing of the $\omega$ and $\phi$ mesons in the natural $q\bar{q}$ basis.

Finally, I show that the ratio of the matrix elements $B_0$ and $B'_0$ for the pseudoscalar and vector mesons is close to one, as expected. By calculating the ratio two ways, I find that overall consistency is obtained for $a\approx -0.18$ and scale factors $x\approx x'\approx 1.05$. The latter parametrize the higher order mass and loop corrections which differ for the $K,\,K^*$ and the $\eta,\,\eta'$ or $\omega,\,\phi$ sectors. The $\eta,\,\eta'$ mixing angle in the usual SU(3) basis is then predicted to be $\theta_P=-15^\circ$. The $\omega,\,\phi$ mixing is small in the natural $q\bar{q}$ basis, $\theta_V=1.95^\circ$, so is nearly ``ideal'' in an SU(3) basis.

I conclude that the whole picture is consistent.

\acknowledgments
This work was supported in part by the U.S. Department of Energy under Grant No.\ DE-FG02-95ER40896.


\input epsf

\end{document}